\documentclass[apj]{emulateapj}
\setkeys{Gin}{width=0.48\textwidth}
\usepackage{graphicx}
\newcommand{\hmpc}{\ifmmode{h^{-1}\,\hbox{Mpc}}\else{$h^{-1}$\thinspace Mpc}\fi}

\newcommand{\kms}{\ifmmode{\,\hbox{km\,s}^{-1}}\else {\rm\,km\,s$^{-1}$}\fi}

\begin{document}
\title{Dynamical Properties of Collisionless Star Streams} 
\shorttitle{Dynamical Properties of Star Streams}
\shortauthors{Carlberg}
\author{R. G. Carlberg}
\affil{Department of Astronomy and Astrophysics, University of Toronto, Toronto, ON M5S~3H4, Canada} \email{carlberg@astro.utoronto.ca }

\begin{abstract}
A sufficiently extended satellite  in the tidal field of a host galaxy loses mass 
to create nearly symmetric leading and trailing tidal 
streams.  
We study the case in which tidal heating drives mass loss from a low mass satellite.
The stream effectively has two dynamical components, a common angular momentum 
core superposed with episodic pulses with a broader angular momentum distribution.
The pulses appear as spurs on the stream,
oscillating above and below the stream centerline, stretching and blurring in configuration space as they move away
from the cluster. Low orbital eccentricity streams are smoother and have less differential motion
than high eccentricity streams. 
The tail of a high eccentricity stream can develop a fan of particles 
which wraps around at apocenter in a shell feature.
We show that scaling the essentially stationary action-angle variables with the cube root of the satellite mass allows 
a low mass satellite stream to accurately predict the features in the stream from a satellite a thousand times more massive.
As a practical astrophysical application, we demonstrate that
narrow gaps in a moderate eccentricity stream, such as GD-1, blur out to
50\% contrast over approximately 6 radial periods. A high eccentricity stream, such as Pal~5,
will blur small gaps in only radial  2 orbits as can be understood from the much larger dispersion 
of angular momentum in the stream.
\end{abstract}

\section{INTRODUCTION}
\nobreak
Stellar streams emanating from dwarf galaxies and globular clusters are of wide interest for their role in the build up of the 
stellar halo of galaxies, and, for their considerable power in mapping the dark matter potential of galaxies
\citep{Dubinski:96,Johnston:96,Johnston:98,Law:05,Binney:08}. 
Density variations along a stream are also useful statistical indicators of
the presence of the substructure within a galaxy halo \citep{Yoon:11,Carlberg:12,Erkal:14} however
their subsequent evolution depends on the dynamical properties of the stream which
are only now beginning to be studied.

Numerical simulations of tidal features drawn out of stellar systems has a long history, 
beginning  with the study of interactions between comparable mass disk galaxies
\citep{Lindblad:61,Pfleiderer:61}.  Increasingly powerful techniques and computer simulations
\citep{CB:72}  demonstrated  that the encounters that produced massive
tidal tails usually led to the galaxies merging \citep{TT:72}.
\citet{Quinn:84} followed the tidal dissolution of low mass disk galaxy infall to produce dramatic tidal streamers and shells
and 
\citet{McGlynn:90} studied the tidal mass loss from small spheroidal systems.
\citet{Kupper:08} showed that streams from almost any stellar system where particles emerge through the tidal lobes 
will create density pile-ups down stream (visible in McGlynn's Fig.~2) 
as a result of the interplay between the epicyclic and orbital motions \citep{Kupper:10,Kupper:12}.

Studies of stellar streams usually assume that the satellite is
collisionless, such as a dwarf galaxy and its dark matter halo, or
that the tidal mass loss process is sufficiently collisionless that 
the  internal dynamics and mass loss from stellar evolution within a globular star cluster can be set aside
and the stream can be adequately realized with a numerical approach using a 
fast solver of a softened gravitational potential rather 
than a direct n-body integrator which carefully
follows the interactions of individual stars \citep{Aarseth:99}.
The basic dynamics of tidal streams has been extensively studied using essentially collisionless methods
useful for exploring many aspects of the galactic potential
 \citep{Johnston:98,Helmi:99,Varghese:11, Kupper:12,Bovy:14,Amorisco:14,Fardal:14}. 
There are wide variations of appearance in  positions and velocities, 
depending on the mass of the satellite, the host potential, 
orbital phase and even the parameters of the n-body model.

Here we are interested in the underlying dynamical details of stellar streams, in particular the distribution of stream 
particles in Hamiltonian action-angle space. 
Action variables for a spherically symmetric potential are simply the angular momentum and 
the radial action, which is proportional to the eccentricity of the orbit. They are conserved quantities.
The angle variables are their canonical conjugates which
evolve linearly in direct proportion to the angular and radial oscillation frequencies of the orbit.
Together these
remarkable dynamical quantities are powerful tools for giving insight into both stream dynamics \citep{EB:11}
and the potential in which the stream is orbiting. 
The isochrone potential is unique in having simple analytic action-angle variables, but
a set of techniques has been developed which allows the remapping of the isochrone into both axisymmetric
and triaxial potentials \citep{McGB:90,MB:08,SB:14, Bovy:14}. 
The distribution of the stream particles in action-angle variables determine the length and width of the streams, 
their offsets from the satellite, and the local internal motions within a stream which are
the quantities of interest for many practical problems.

In this paper we use a standard n-body tree code 
 to  evolve a low mass King model satellite in an isochrone potential.  
The distribution of stream particles in action-angle space is derived and related to the evolution
of the tidal stream and its features. We test for the expected scaling with tidal radius for satellites 
of differing mass on the same orbit.
A particular application is to follow the evolution
of narrow gaps in a stellar stream before they become hard to detect.
The overall goal of this paper and others studying stream dynamics in simplified host galaxy potentials
is to build up a dynamical framework 
that is useful for understanding stream dynamics in the irregular and evolving potentials of real galaxies.

\section{N-body Simulations}

We are interested in the tidal dissolution of a very small galaxy or star cluster
 in the potential of the host galaxy.  Accordingly we place a small n-body system representing 
the satellite 
within a fixed external potential, $\Phi(r)$, representing the host, and assume that the satellite mass
is low enough that dynamical friction can be neglected.
The satellite is a low concentration King model with a central potential, $\phi_0$, relative
to the central velocity dispersion, $\sigma$, of  $\phi_0/\sigma^2=-4$. 
More concentrated satellites can be easily generated, but not surprisingly, their low outer densities
reduce   mass loss, making for thinly populated, but otherwise identical streams.
The King model satellite is assigned a mass, $m$, and scaled to have an outer radius equal to the formal 
 tidal radius, $r_J=r[m/(3M(r))]^{1/3}$ for a Keplerian potential.
The expression is evaluated at the initial orbital radius $r$ (which here is the apocenter of the orbit) 
where the host mass interior to $r$ is $M(r)$. 
The correction to a non-Keplerian potential is,
 $(1-{1\over 3}{d\ln{M(r)}/{d\ln{r}}})^{1/3}$,
\citep{BT:08}, which at a location with a flat rotation curve is, $(2/3)^{1/3}$, a reduction 
of the tidal radius of  13\%. We do not normally include this factor
since it reduces the already low mass loss rate, and, otherwise makes little visible difference to the stream.

The self-gravitating satellite orbits in an external potential which represents a host galaxy.
The satellite is launched from apocenter, that is the initial velocity is purely tangential, at some fraction
of the local circular velocity or angular momentum, $L/L_c$. 
We use the isochrone potential, $\Phi(r)=-GM/(b+\sqrt{b^2+r^2})$, where $M$ is the total mass and $b$ the scale radius.
We chose units in which the isochrone potential is assigned unit mass and has unit scale radius. 
The satellite mass then is a ratio
to the total mass of the system, with our standard value being $10^{-7}$. The resulting 
satellite corresponds to a low mass, low concentration globular cluster.

We start the satellite at $r=2.2$, which is just outside the maximum of the circular velocity, 
$r=\sqrt{2\sqrt{2}+2}\simeq 2.197$, of the isochrone. 
The isochrone has a large core which makes relatively elliptical orbits easy to handle but leads 
to relatively low mass loss for our chosen apocenter. 
Satellites started at larger radii with the same angular momentum relative to a circular
orbit have substantially more mass loss because they do not orbit through the core.

\begin{figure}
\begin{center}
\includegraphics[angle=0, scale=0.9]{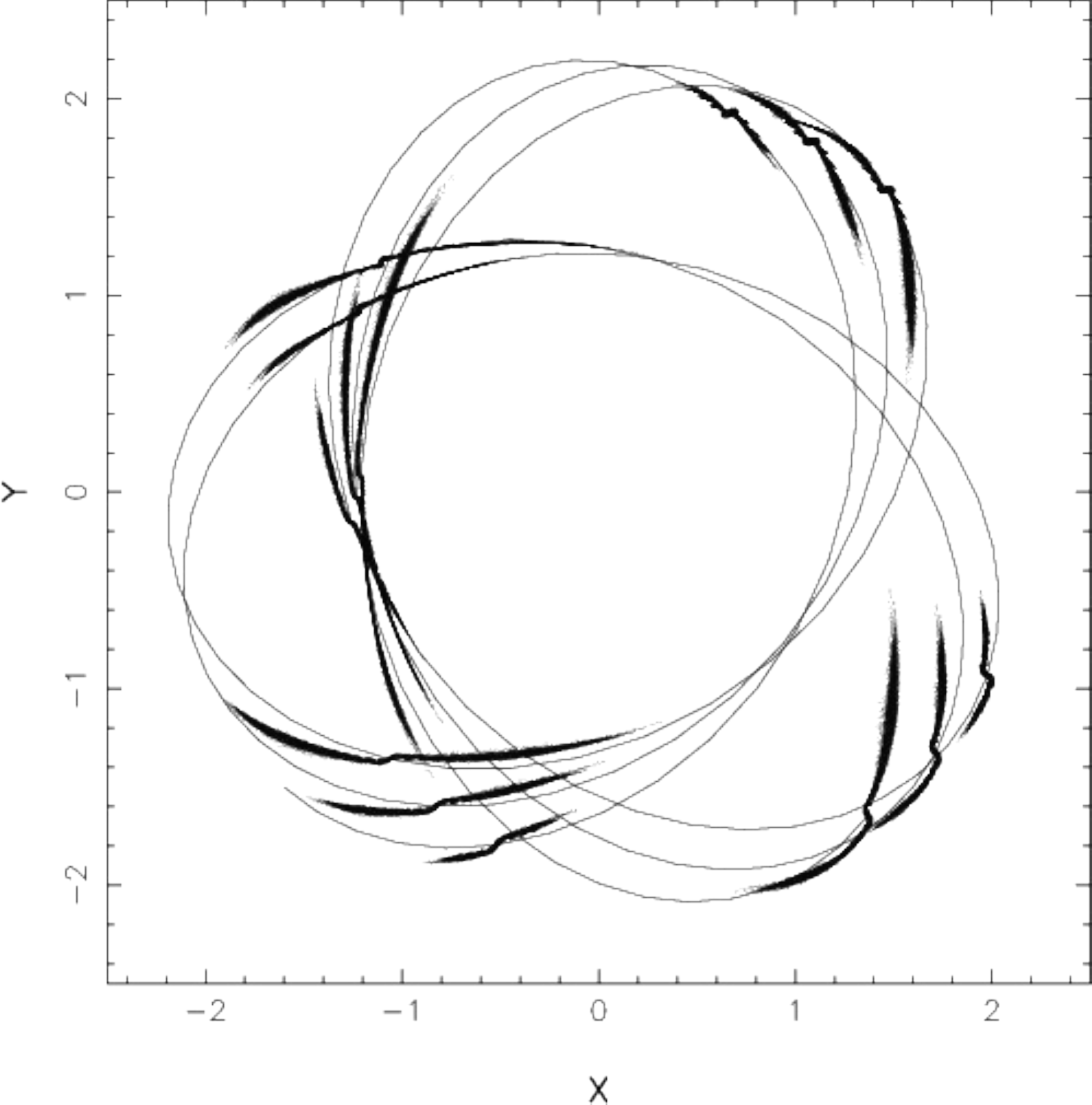}
\end{center}
\caption{The development of  tidal tails for an $L/L_c=0.7$ orbit. The stream is displayed 
at intervals of 10.9 time units, from time 73.0 to 215.3. 
 The satellite is orbiting counter-clockwise along the thin line.
}
\label{fig_xy}
\end{figure}

\begin{figure}
\begin{center}
\includegraphics[angle=-90, scale=0.7]{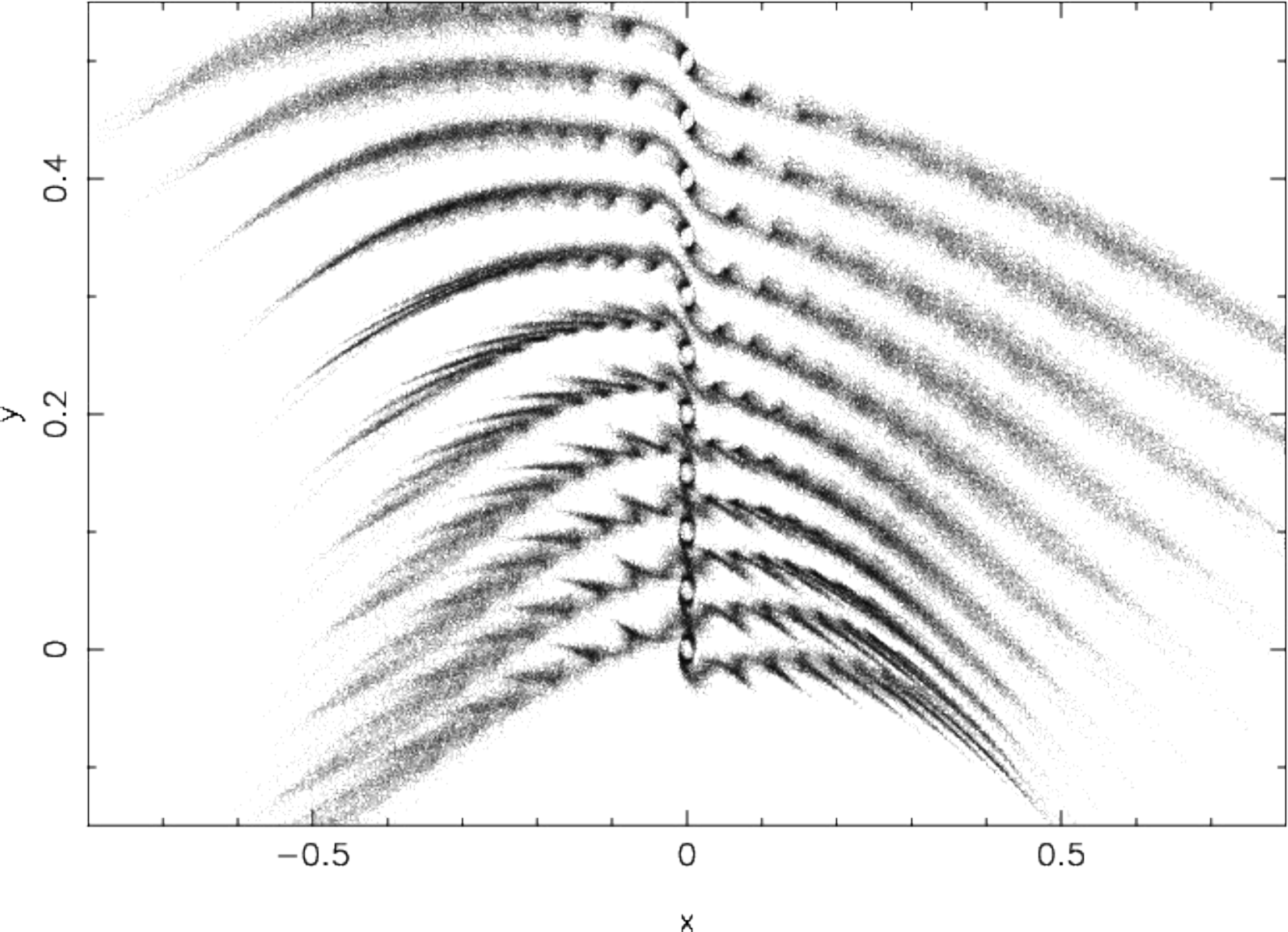}
\end{center}
\caption{The development of the $L/L_c=0.7$ stream from time 160.6 (bottom) to 167.9 (top)
at intervals of 0.73 in time which straddles the orbital apocenter.
The satellite is located at $x=0$ with the y axis shifted upward 0.05 
units for each plot. The satellite is moving along the x axis, which is measured in linear
units along the stream, to the right in all cases. The vertical
scale here is stretched 14 times relative to the horizontal scale. Points inside the tidal radius of 0.01  are not plotted.
}
\label{fig_xylocal}
\end{figure}

All the simulations analyzed in detail here start at apocenter $r_a=2.2$.
For  clusters on orbits  with angular momentum relative to the circular value, $L/L_c$,  of 0.4, 0.7 and 0.9, the 
pericenter radii, $r_p$, are 0.604, 1.214 and 1.806, respectively.  For eccentricity defined as $e=(r_a-r_p)/(r_a+r_p)$,
the orbits have eccentricities of  0.57, 0.29 and 0.10 respectively.
The related $(\Omega_\phi, \Omega_r)$ 
pairs are respectively (0.246,0.417), (0.231,0.355), and (0.206,0.298) with the two oscillation periods being $T=2\pi/\Omega$, 
which are in the neighborhood of 20 time units.

\begin{figure}
\begin{center}
\includegraphics[angle=-90, scale=0.8]{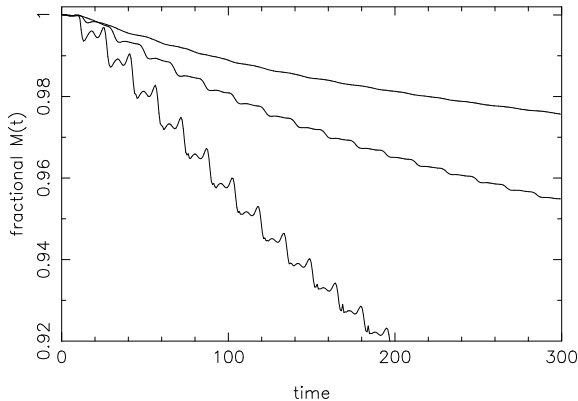}
\end{center}
\caption{Mass within 2 tidal radii defined at the starting time for orbits with $L/Lc$ of 0.4, 0.7,
and 0.9 from bottom to top, respectively.
}
\label{fig_mt}
\end{figure}

The host potential is fixed but the self-gravity of the satellite system is 
calculated dynamically, which is important to accurately follow the creation of the streams.
Our reference simulations use the well-known Gadget2 code \citep{Springel:05}
modified to include an external fixed galactic potential, as described in \citet{NC:14}. 
We normally start with $10^6$ particles in the satellite.
The code parameters are largely set at standard values, with 
a  softening of 1/3000 of the scale radius (5pc if scaled to a 15 kpc isochrone scale radius),
approximately 24\% of the core radius of the n-body cluster. 
The time integration error tolerance is set to 0.015 and the force error to 0.003. 
Once well outside the cluster the angular momentum is conserved to a fractional accuracy of $10^{-5}$ over eight radial periods.

For very low mass satellites energy and angular momentum conservation checks are
not particularly good tools for monitoring the accuracy of the calculation. 
Our approach is to repeat  the run with modified error control parameters and compare the distribution of stream particle positions. 
Runs with the time integration error tolerance reduced to 0.0075 and increased to 0.025 relative to the
standard 0.015 value are done to test the stability of the distribution of particles in the streams.
Within the cluster the orbits are effectively chaotic, but we want to be sure that the statistical distributions in the streams are stable. 
After about 3 orbits the mean stream particle angular momentum offset from the satellite is identical in all cases
within 1\% and the dispersion about the mean is only different at 1\%, with a handful of particles becoming outliers.
The calculation with the larger time step error control appears to be nearly statistically identical with the more
tightly controlled simulation
but we use the intermediate time step error control because it is certainly statistically stable
to the limits we require.

\section{The Development of Tidal Tails}

The orbital angular momentum of the satellite determines the orbital eccentricity, which
is a key factor in the dynamics of the tidal tails. Both cosmological simulations and observed
streams guide our choice of orbits to consider. The dark matter sub-halo orbital angular
momentum distribution from cosmological simulations (Benson 2005; Wetzel 2011; Jiang et
al. 2014) has a median $L/L_c$ of about 0.5, with a median orbital eccentricity of about 0.7.
\citep{Odenkirchen:03,Dehnen:04} and  0.33 \citep{Willett:09}, respectively.  We will do simulations
close to both these orbital eccentricities, although within the spherical isochrone potential.

In Figure~\ref{fig_xy} we show the x-y projection of the stream and satellite in an $L/L_c=0.7$ orbit at 14
times spaced 10.9  time units apart over 8.04 radial periods. The orbit of the center of the satellite is shown as the thin line.
As is well known the stream is usually significantly offset from the orbit of the satellite.
In Figure~\ref{fig_xylocal} we use the satellite as the local coordinate origin and the direction of motion as the positive $x$ axis to
show the stream in the frame of the satellite over 6.3 units of time, which is about 1/3 of an orbit.  
The stream is plotted at multiple times to show its structural evolution.
Note that time runs upwards.
The stream has a core along with a set of spurs \citep{Dehnen:04,Amorisco:14,Fardal:14,Hozumi:14}
 which accompany each pulse of mass loss which begins with heating at pericenter and particles drifting away at apocenter.
The spurs oscillate above and below the core of the stream and gradually lengthen and blur 
together.

The mass inside $2r_J$ as a function of time is shown in Figure~\ref{fig_mt} for various models
with $m=10^{-7}$. 
The rate of mass loss in an isochrone potential is fairly low, 5\% over 300 time units, or, about 0.3\% per radial orbit for 
our standard satellite on an $L/L_c=0.7$ orbit. 
The low mass loss is a consequence of the isochrone's large core radius, inside of which
the tidal forces are low and the nearly harmonic potential means that particles do not stream away
from the satellite.  Moving the orbital apocenter to $r=3$ for an $L/L_c=0.7$ orbit
nearly doubles the mass loss rate relative to an $r=2.2$ orbit as shown in Figure~\ref{fig_mt}.
The satellite on an $L/L_c=0.4$ orbit has a mass loss rate double that for the same satellite
on an $L/L_c=0.7$ orbit. Reducing the outer ``tidal" radius of the initial satellite by 10\% approximately halves the mass
loss rate.  Most of the properties of the stream are set by the satellite mass and the satellite 
orbit. The absolute value of the mass loss rate is not usually important, until the mass loss is so high that the satellite mass
changes substantially, which we do not consider here.

\section{Stream Particle Action-Angle Variables}

Action-angle variables have been widely discussed as a powerful way to examine
the dynamics of streams \citep{EB:11,Bovy:14}. 
The actions and angles are calculated  (up to factors of $2\pi$) from the instantaneous 
positions and velocities using the isochrone formulae given in \citet{BT:08} (the sign error in the $J_r$ equation is noted on 
the authors' website). In a spherical potential all orbits are confined to planes.
For the thin tidal tails explored here we use the total angular momentum 
rather than just the component around the perpendicular to the orbital plane of the satellite, $L_z$. 
The two differ for each particle by only a small, constant, tilt factor, roughly the tidal radius relative to the orbital radius, 
which is about $10^{-2}$ here.

\begin{figure}
\begin{center}
\includegraphics[angle=-90, scale=0.8]{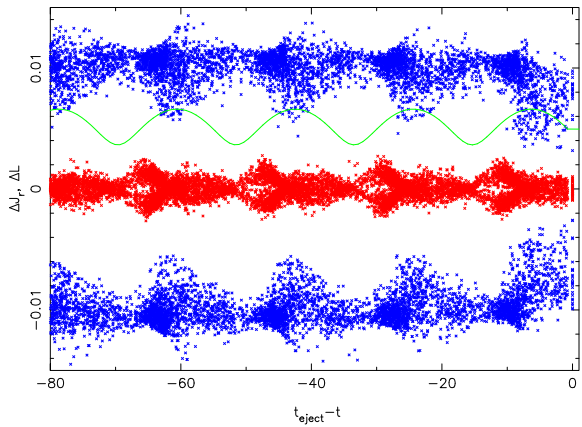}
\end{center}
\caption{
The time 242 values of $\Delta J_r$ (inner, red, scaled to 0.01) and $\Delta L$ (outer, blue, scaled to 0.015) measured
relative to the satellite as a function
of ejection time. 
The actions are calculated with respect to the center of the satellite system. 
The isochrone's actions are not conserved near the satellite.
The (green) line is proportional to the radial position of the satellite.
 }
\label{fig_jrpt}
\end{figure}

In doing this calculation we make the approximation that once particles leave the satellite its potential can be ignored. 
The error
is roughly $(r/d) m/M(r)$, where $m$ is the factional mass in the satellite and $d$ is the distance of the particle from
the satelllite. For our standard satellite mass of $m=10^{-7}$, once $d$ becomes about 10 tidal radii the effects of the satellite are negligible.
The gravitational effect of the satellite is visible in the first ten units of time in
the action plots below, Figures~\ref{fig_jrpt}, \ref{fig_jrpt34}, and \ref{fig_jrpt39}.

\begin{figure}
\begin{center}
\includegraphics[angle=-90, scale=0.8]{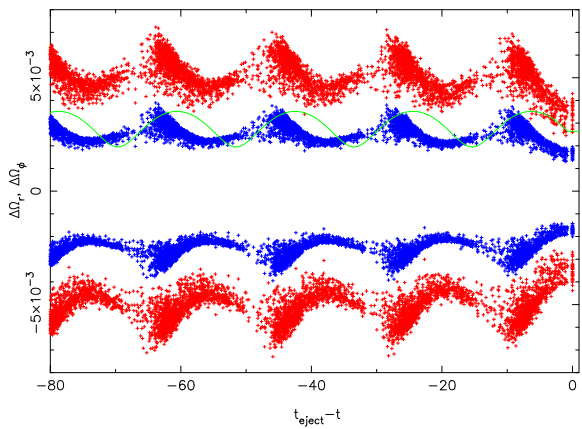}
\end{center}
\caption{The frequencies $\Delta \Omega_r$ (red) and $\Delta \Omega_\phi$ (blue) at time 242 relative
to the frequencies of motions of the center of the satellite are plotted against age in stream
for an $L/L_c=0.7$ orbit. Both frequencies are scaled to a maximum of 0.06.
The (green) line is proportional to the radial position of the satellite.
}
\label{fig_omrpt}
\end{figure}

\begin{figure}
\begin{center}
\includegraphics[angle=-90, scale=0.8]{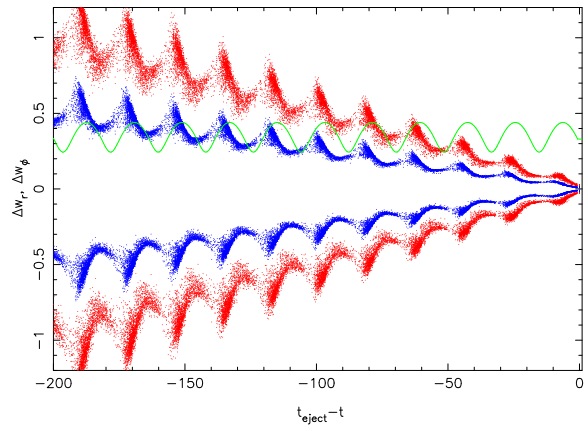}
\end{center}
\caption{The angles $\Delta w_r$ (red) and $\Delta w_\phi$ (blue) relative to the satellite 
at time 242  are plotted against age in stream
for an $L/L_c=0.7$ orbit.
The (green) line is  proportional to the radial position of the satellite. Note that the time
duration is longer than in Figures~\ref{fig_jrpt} and \ref{fig_omrpt}.
}
\label{fig_wrpt}
\end{figure}

\subsection{Dynamics of an $L/L_c=0.7$ Stream}

We measure the  stream actions, $\Delta L$, $\Delta J_r$, and angles, $\Delta w_\phi$, $\Delta w_r$,
 relative to those of the center of the satellite. 
Figure~\ref{fig_jrpt} shows the relative actions evaluated at time 242, then displayed as a function of the time that the particles
have been in the stream, $t_{eject}-t$.
The ejection time is calculated as the time at which the
particle crossed the $2r_J$ surface. 
The plot shows pulses of particles emerging through the $2r_J$ surface near apocenter after
 tidal heating around pericenter.  
For a low mass satellite the inner and outer tidal tails are essentially symmetric relative to the satellite. 
The angular momentum of the main pulse of particles is concentrated around a constant mean, with a tail mainly
towards lower values. 
Within each pulse of mass loss $\Delta J_r$ rises to a maximum with ejection time. 
Since the satellite mass does not change
much with time, the pattern of $\Delta J_r(t)$ and $\Delta L(t)$ repeats every orbit.

Once the particles are in the stream the angles evolve 
at a rate fixed by their angular frequency, $w_r=\Omega_r t$.
In general, particles with higher angular momenta relative to the satellite will have higher frequencies,
as shown in Figure~\ref{fig_omrpt}. The range of frequencies means that the particles will
spread with time and eventually start to overlap other bunches of particles, which
is clearly shown in the angles displayed in Figure~\ref{fig_wrpt}.
An important feature is that $\Delta\Omega_\phi$ is about half of $\Delta\Omega_r$. 
Therefore,  the  differential
angular momentum only slowly mixes particles into the same physical regions.
On the other hand the radial action of the pulse has a gradient with time
which leads to the spurs on the stream, as 
shown in Figure~\ref{fig_xylocal}. 
The spurs oscillate up and down relative to the centerline of the stream. The spurs also stretch
with time since the $\Omega_r$ and $\Omega_\phi$ are strongly correlated in the pulse of
particles.  

\begin{figure}
\begin{center}
\includegraphics[angle=-90, scale=0.85]{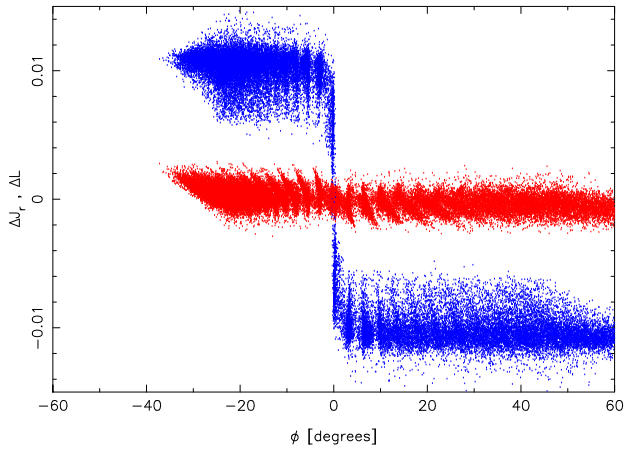}
\end{center}
\caption{
An illustration of how the action variables project into configuration space.
$\Delta J_r$ (central points, red) and $\Delta L$ (high and low points, blue) measured relative to the
satellite center  versus the azimuthal angle of the stream at time 242
for an $L/L_c=0.7$ orbit.  The asymmetry with respect to the satellite
varies with orbital phase.
 }
\label{fig_jrp_phi}
\end{figure}

Figure~\ref{fig_jrp_phi} shows the distribution of $\Delta J_r$ and $\Delta L$ as a function of $\phi$, the azimuthal angle 
along the stream as seen from the galactic center.
The plot is made at time 242. The distribution shows the expected sorting particles with
large 
angular momentum differences move relatively further down the stream. 
The sorting helps the stream phase mix with distance from the satellite, but the stream 
never has a simple uniform distribution in the actions, and, the distribution varies 
down the length of the stream.

\subsection{Dynamics of  $L/L_c$ 0.4 and 0.9 Streams}

\begin{figure}
\begin{center}
\includegraphics[angle=-90, scale=0.8]{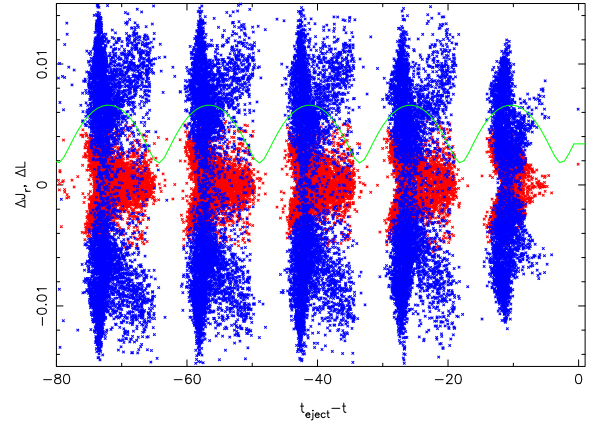}
\end{center}
\caption{
The time 242 values of $\Delta J_r$ (inner, red, scaled to 0.01) and $\Delta L$ (outer, blue, scaled to 0.015) measured
relative to the satellite as a function
of ejection time for the $L/L_c=0.4$ stream.
The actions are calculated with respect to the center of the satellite system.
The (green) line is proportional to the radial position of the satellite.
 }
\label{fig_jrpt34}
\end{figure}

\begin{figure}
\begin{center}
\includegraphics[angle=-90, scale=0.7]{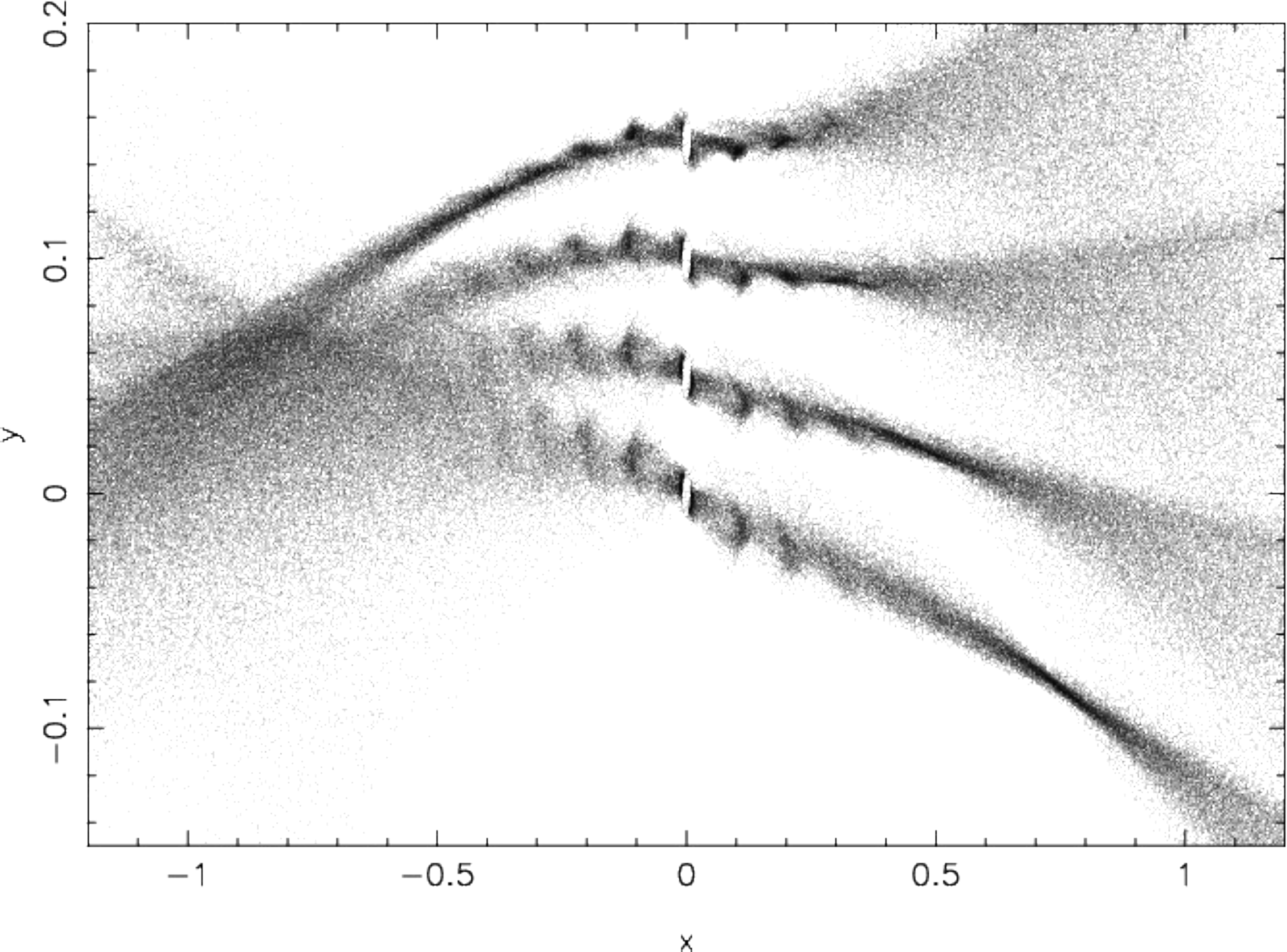}
\end{center}
\caption{The development of the $L/L_c=0.4$ stream from time 238.0 (bottom) to 240.2 (top)
at intervals of 0.73 in time which straddles the orbital apocenter. The axes are the same
as in Figure~\ref{fig_xylocal}, although with a different scale.
}
\label{fig_xyl34}
\end{figure}

\begin{figure}
\begin{center}
\includegraphics[angle=-90, scale=0.9]{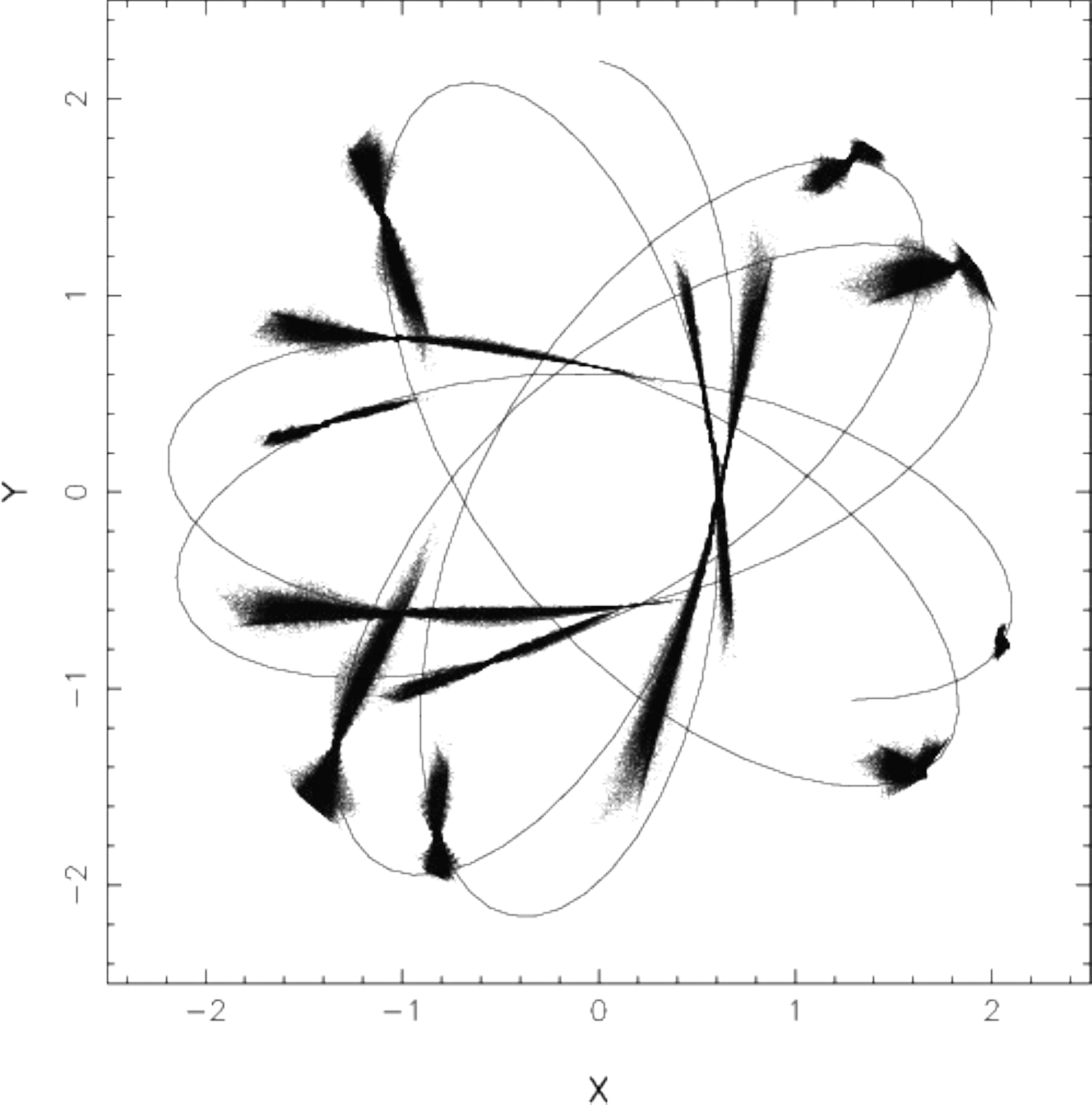}
\end{center}
\caption{The $L/L_c=0.4$ stream in the x-y plane for the same
times as Figure~\ref{fig_xy}.
}
\label{fig_xy34}
\end{figure}

A cluster on a high eccentricity orbit will have strong tidal heating at the pericenter, hence mass loss
will be more concentrated in time,  as is shown in Figure~\ref{fig_jrpt34}. 
The angular momentum distribution of the ejected material covers a broader
range than for the lower eccentricity orbit of Figure~\ref{fig_jrpt}.
The resulting tidal stream has very strong shear that rapidly blurs out all features, Figure~\ref{fig_xyl34}.
The stream is shown as it passes pericenter, when it becomes thinner. 
The fan-like feature at the end of the stream is 
the result of local orbital mixing producing a fairly smooth distribution in actions near the end of the stream, and
the fact that the end of the stream is near apocenter, so that the material is spread out perpendicular to the stream
as shown in Figure~\ref{fig_xy34}. A stream on a highly eccentric orbit
would be relatively hard to detect except for the portion of the stream which happens to be near pericenter.

\begin{figure}
\begin{center}
\includegraphics[angle=-90, scale=0.8]{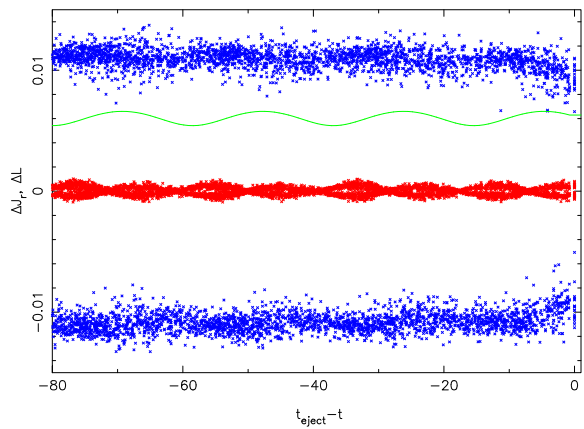}
\end{center}
\caption{
The time 242 values of $\Delta J_r$ (inner, red) and $\Delta L$ (outer, blue) measured
relative to the satellite as a function
of ejection time for the $L/L_c=0.9$ stream.
The actions are calculated with respect to the center of the satellite system.
The (green) line is proportional to the radial position of the satellite.
 }
\label{fig_jrpt39}
\end{figure}

\begin{figure}
\begin{center}
\includegraphics[angle=-90, scale=0.7]{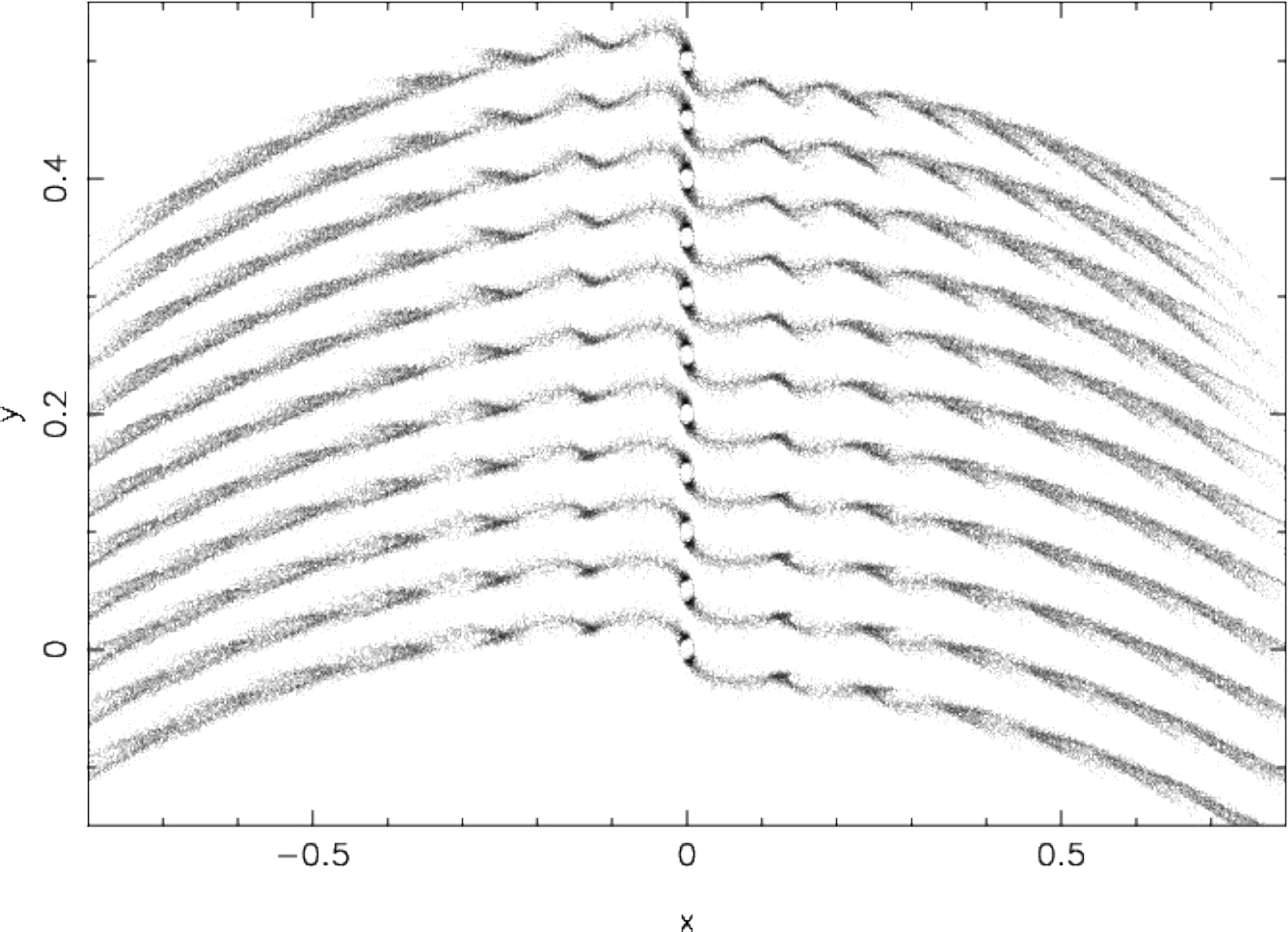}
\end{center}
\caption{The development of the $L/L_c=0.9$ stream over precisely
the same time range as shown for the $L/L_c=0.7$ stream in Figure~\ref{fig_xylocal}.
}
\label{fig_xyl39}
\end{figure}

A cluster on a nearly circular orbit, $L/L_c=0.9$, has very mild tidal heating. The outcome 
is nearly continuous mass loss with relatively little dispersion in the angular momentum and
relatively small radial actions as shown in Figure~\ref{fig_jrpt39}. The result is a stream with relatively simple behavior, 
however it retains the same basic features as the other streams, with the spurs  still present, 
as shown in Figure~\ref{fig_xyl39}.

\subsection{Stream Dynamics variations with $L/L_c$}

\begin{figure}
\begin{center}
\includegraphics[angle=-90, scale=0.8]{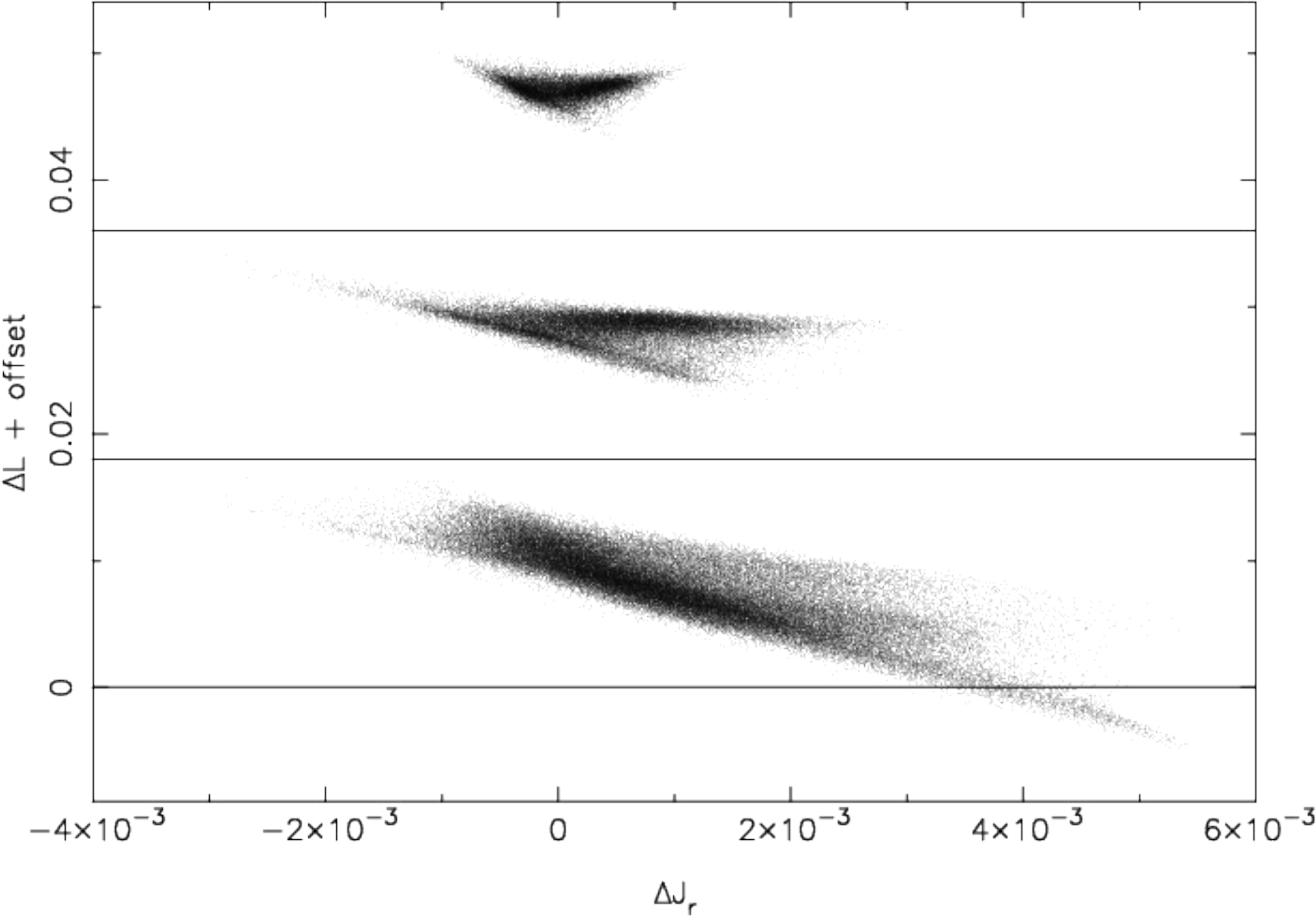}
\end{center}
\caption{The distribution of actions, $\Delta J_r, \Delta L$, relative to the satellite.
The orbits have  $L/L_c$  0.9, 0.7 and 0.4, from top to bottom, respectively,
  all evaluated at time 219, although these distributions are very stable with time.
Each plot is offset 0.018 vertically. The lines indicate $\Delta L=0$ for each set of points. 
}
\label{fig_jrp3d}
\end{figure}

In Figure~\ref{fig_jrp3d} the joint distributions of stream particles 
in the actions, $\Delta J_r$ and $\Delta L$, relative to the satellite are shown for the three
satellite orbits, $L/L_c$ of 0.9, 0.7 and 0.4.  The inner and outer stream values are essentially completely
symmetric so have overplotted, with the inner
stream values reflected through the origin.
The  mean angular momentum offset of the stream decreases from 0.0111, to 0.0102, to 0.0079 with the dispersion
increasing from 0.0008, to 0.0012 to 0.0029, as the
satellite's angular momentum declines from an (absolute) value of 2.312, to 1.837, to 1.415, respectively.
In $\Delta J_r$ the most nearly circular orbit  is a nearly symmetric distribution, similar to that seen in \citet{Dehnen:04,EB:11,Bovy:14}.
As the eccentricity of the orbit grows both the overall spread of the distribution and its asymmetry grow as a
consequence of the increasing size of the pulse of mass loss which produces a correlated $\Delta L-\Delta J_r$ distribution.
Although streams are not very well represented by mean quantities and dispersions about them, their
changes with orbital shape does provide useful guidance for trends in the dynamics of the streams.

\section{Mass Scaling of Streams}

\begin{figure}
\begin{center}
\includegraphics[angle=0, scale=0.9]{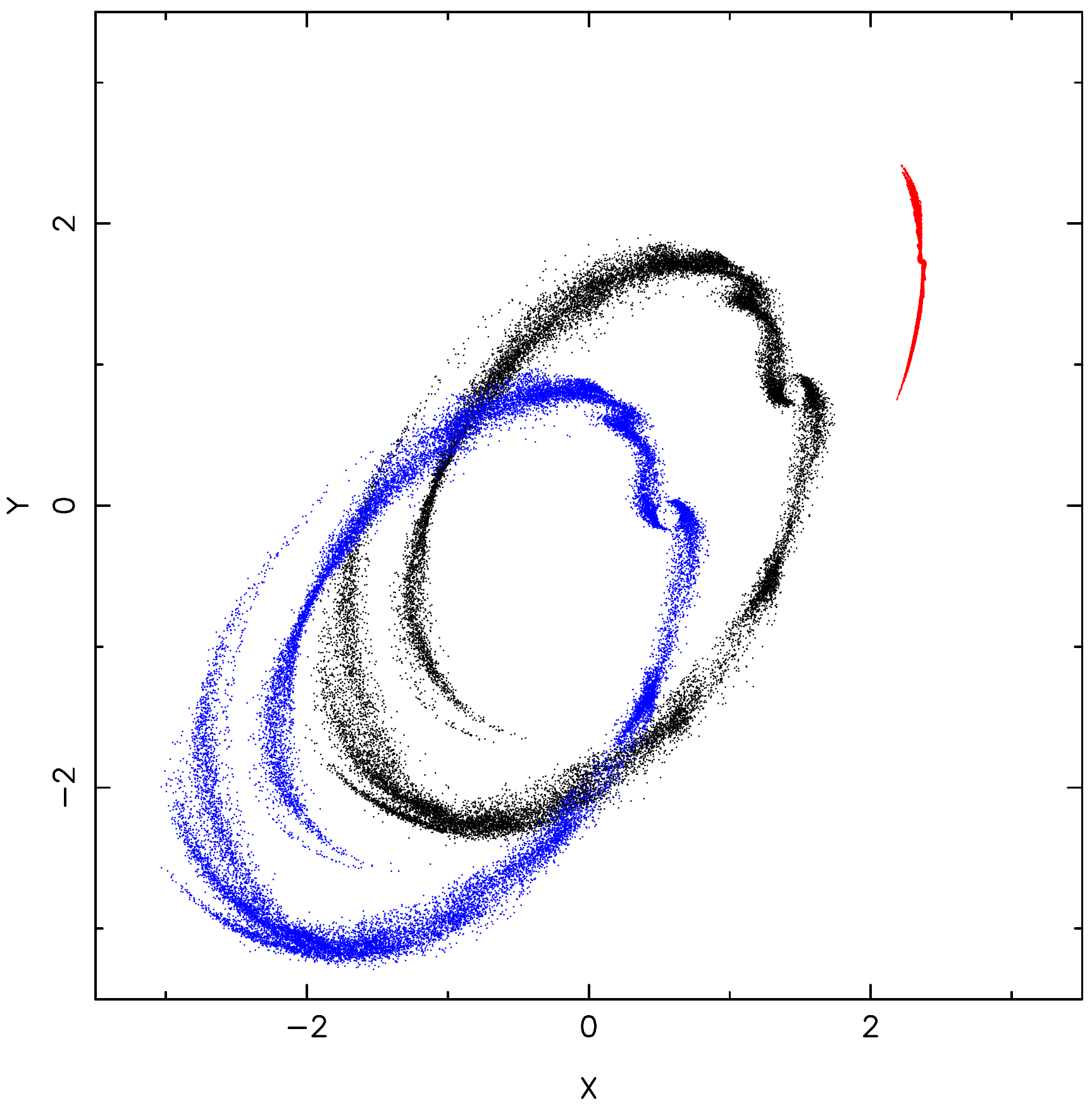}
\end{center}
\caption{A satellite of mass $10^{-7}$ (red points)  scaled to mass $10^{-4}$  (black points
centered on the origin)
using the cube root of the mass ratio with the actions and angles at time 139.4.
A completely independent n-body simulation of a $10^{-4}$ mass satellite is shown offset to the lower left (blue points)
to demonstrate the accuracy of the prediction.
 }
\label{fig_xy_heavy}
\end{figure}

The properties of streams should scale with the tidal radius of the satellite, that is, as the 
cube root of the satellite mass as fraction of the halo mass, for satellites on the same orbit \citep{SB:13}.
Although streams emanating from more massive satellites are longer and wider than those from lower mass satellites
the distributions of actions and angles are scaled versions of each other. 
The scaling works as follows. First calculate the actions and angles for a reference stream, 
apply the $m^{1/3}$ factor to all the $J$ and $w$ variables, then, calculate the positions and velocities
from the new actions and angles. 
We show how well the idea works over the impressively large mass range of streams
from  $m=10^{-7}$  and  $m=10^{-4}$ satellites. 
Figure~\ref{fig_xy_heavy} shows the results.
The very small stream from the  $m=10^{-7}$ satellite is located in the upper right (red points) and the
much larger stream from a $m=10^{-4}$ satellite, done in a completely independent simulation,
 is offset to the lower left (blue points).  
The black points are the scaled up version of the low mass stream and is a very good
match to the n-body simulation for the heavier satellite.
In this particular case the stream is sufficiently short and well positioned in azimuthal
angle that there are no ambiguities of $2\pi$ in the scaled variables. At later times the scaling requires using the time of ejection to calculate the
angles and keeping track of the $2\pi$ wrapping. 

The accuracy of this scaling means that once a low mass stream has been calculated for a given orbit it can be scaled to any other 
satellite mass. The tidal tails of the low mass system might appear denser in Figure~\ref{fig_xy_heavy}, but the scaling means 
that the physical densities are identical, independent of mass. Of course the phase density is much higher in the low mass stream (1000 times here)  and that the column density is higher in the high mass stream (10 times here).

Although these simulations were oriented towards visible star
clusters and dwarf spheroidal galaxies, they are also applicable to the dissolution of pure dark matter sub-halos.
The substantial internal structure in the streams means that 
should a stream pass through a direct dark matter
detector on earth that the detection rate  will vary with location in the stream roughly in proportion to the local density
\citep{LewinSmith:96}. The local density
along the stream varies  by  a factor of order unity, due to the orbital variations in the mass loss rate and
features such as the shells  in Figure~\ref{fig_xy34} \citep{Tremaine:99}. However, the density in the stream  
is generally of order one percent or less of the background density.
That is, if there is as much as  10\% mass loss per radial orbit, with the mass spread over a volume 
oft length of  $2\sqrt{2}\pi\simeq 10$ satellite radii with a half width comparable to the satellite radius, 
with the mean density inside the satellite 
radius being comparable to the local density of the background halo,
hence the mean stream density relative to the local background is
 100 times lower than the background dark matter density \citep{Vogelsberger:09} so
the stream density variations will only modulate the small extra component from the stream.
Orbits where the stream reaches a region of relatively low local dark matter density 
and has increased interactions at low velocities can produce larger effects \citep{Sanderson:12}.

\section{Gap Blurring in Streams}

One of the interests of stellar streams is that they will develop density variations along
the stream in response 
to the myriad sub-halos predicted \citep{VL1,Aquarius} to be orbiting within a 
galactic halo \citep{Carlberg:12}.
The gaps will be blurred out 
at a rate which depends on the size of the gap and 
the distribution of the stream particles in action-angle space.

\begin{figure}
\begin{center}
\includegraphics[angle=-90, scale=0.7]{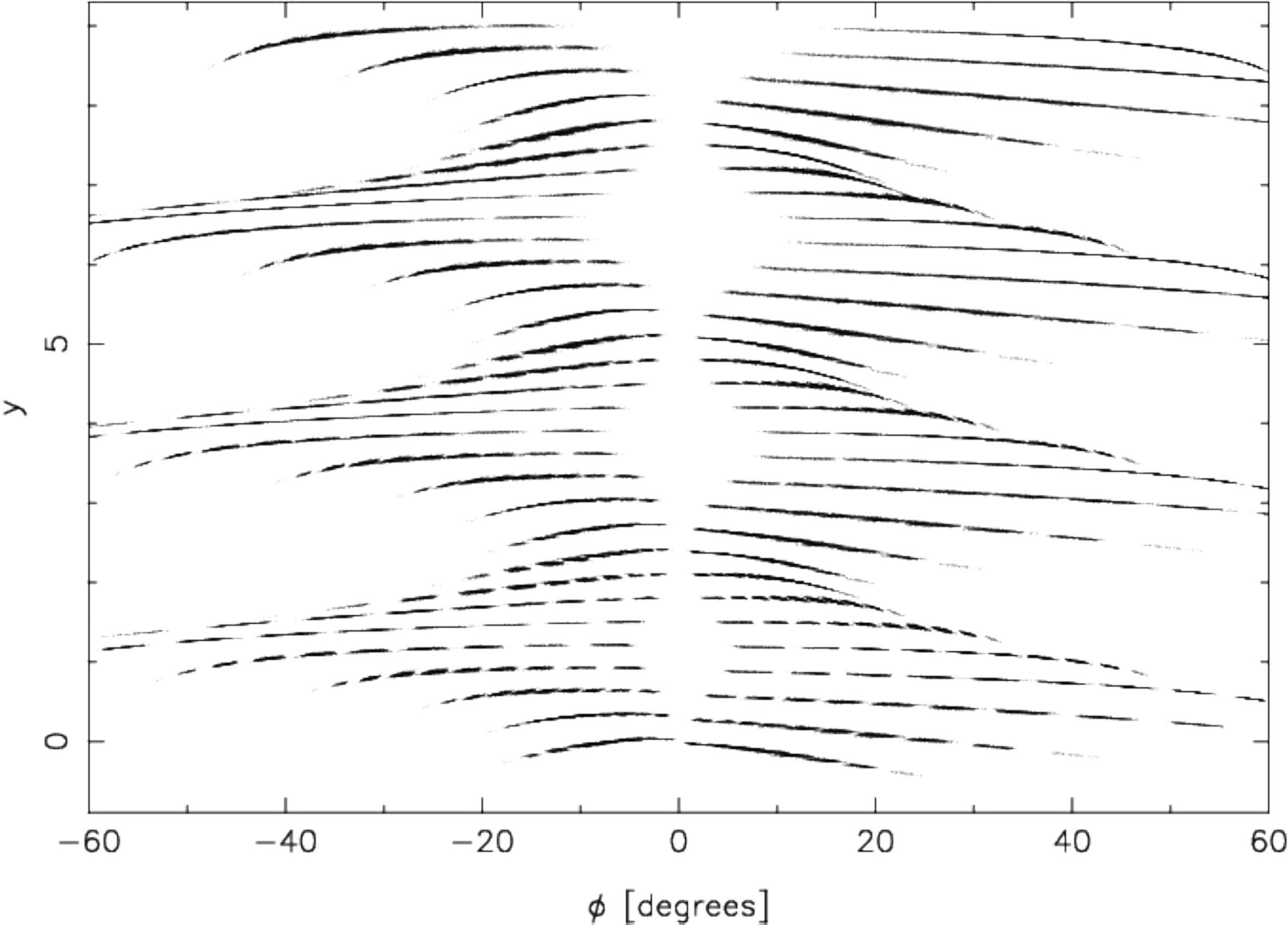}
\end{center}
\caption{Evolution of artificially inserted gaps of 3 degrees every 6 degrees inserted into the $L/L_c=0.7$ stream near pericenter.}
\label{fig_xylholes}
\end{figure}

\begin{figure}
\begin{center}
\includegraphics[angle=0, scale=1]{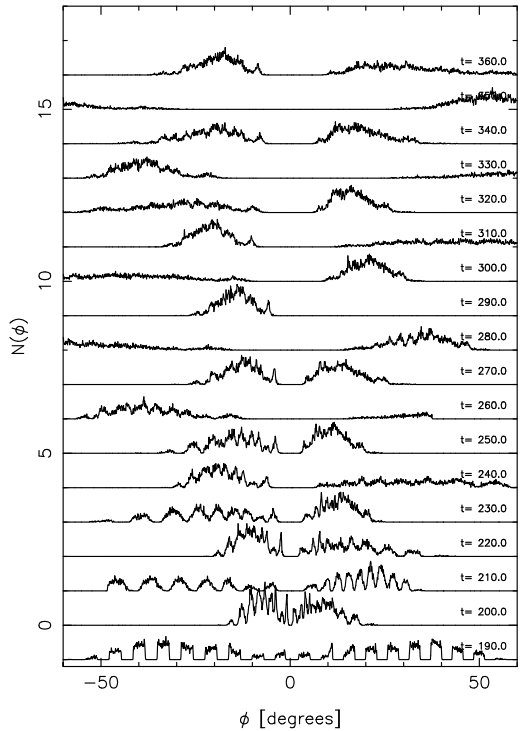}
\end{center}
\caption{Evolution of artificially inserted gaps of 3 degrees every 6 degrees inserted into the $L/L_c=0.7$ stream near pericenter.
The radial period is 17.7 time units and the azimuthal period is 27.2 time units. Simulation times are
indicated on the plot.
}
\label{fig_holes7}
\end{figure}

To create a stream for which we can easily measure that
rate at which gaps blur out we take one of our calculated streams and simply remove particles 
along the stream to create gaps. We use a time
 shortly beyond pericenter, time 190 at $r=1.23$ 
in the $L/L_c=0.7$ simulation with $10^6$ particles. We want to determine how long gaps about
as wide as the stream last. 
Yet wider gaps will last longer in direct proportion to their width.
In the stream near pericenter we
remove all particles in
gaps of 3 degrees wide every 6 degrees and remove the satellite itself, since it has almost no 
effect on the stream particles, and then
evolve the stream forward in time. 
The stream with the added gaps is shown in x-y in Figure~\ref{fig_xylholes} and then
as a linear density measurement of density along the stream in angular coordinates in Figure~\ref{fig_holes7}.
The linear gap width depends on location, but is comparable to the width of the stream.
We do the same for the stream as it passes around apocenter, finding sufficiently similar behavior that
we do not display it.

Figure~\ref{fig_xylholes} shows that epicyclic motions tilt the 
gaps back and forth \citep{Yoon:11} which leads to an apparent blurring of the density when viewed 
from the center of the galaxy in angular coordinates, Figure~\ref{fig_holes7}. It also means that the visibility 
of gaps will depend on viewing angle. The illustrative density profiles here view the streams from the center of the galaxy.
There is an accompanying slow decline as a result
of differential motions of the material across the stream which fill in the gaps.
This $L/L_c=0.7$ orbit has a radial period of 17.7 time units. Although the
visibility varies with orbital phase, the gaps steadily decline in contrast to about 50\% amplitude, roughly the limit of detectability
in the low signal-to-noise filtered sky images that are available,  after about 100 times units, approximately 6 radial orbits.
Scaling to an orbit with a radial period
of about 1\,Gyr, we conclude that narrow gaps in a moderate eccentricity stream, like GD-1, would be visible for a substantial 
fraction of the age of the galaxy. 

\begin{figure}
\begin{center}
\includegraphics[angle=0, scale=1]{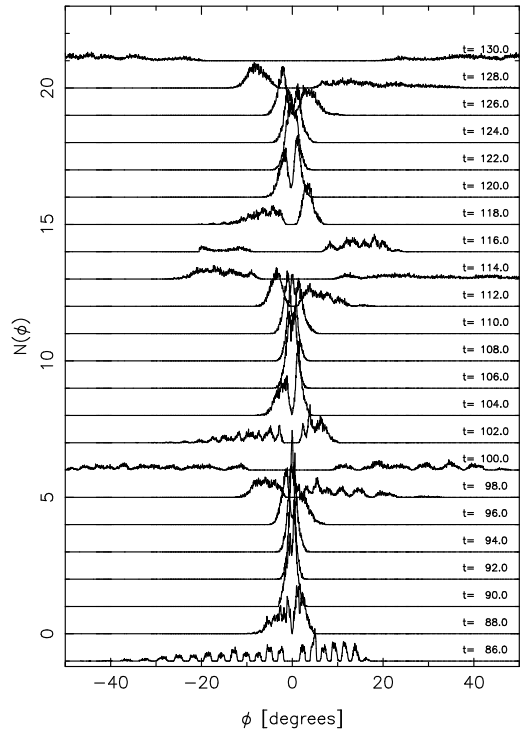}
\end{center}
\caption{Evolution of gaps of 2 degrees every 4 degrees inserted into the $L/L_c=0.4$ stream near pericenter.
Note that the time scale is much finer than in Fig.~\ref{fig_holes7}.
}
\label{fig_holes4}
\end{figure}

Gaps of 2 degrees inserted every 4 degrees  at time 85.4 in the high eccentricity $L/L_c=0.4$ stream as it orbits
near pericenter is shown in
Figure~\ref{fig_holes4}. The visibility of the gaps again depends on orbital phase, but they remain visible for 30 time units,
where the radial period is 15.1 units, 
but then disappear. 
Therefore narrow gaps in the high eccentricity stream Pal~5, will likely only  last two orbits,
as expected from the roughly 2.5 times higher dispersion of the stream angular momentum, Figure~\ref{fig_jrp3d}.
A better estimate requires integrating the orbit in a more accurate potential model of the Galaxy. 
Only the narrow gaps are removed quickly since 
the duration that a gap is visible is proportional to its width, so wider gaps will survive proportionally longer.

\section{Conclusions}

We have simulated the tidal dissolution of self-gravitating satellites 
primarily to study 
the dynamical properties of the tidal tails. 
Mass loss 
is a consequence of tidal heating  in these collisionless simulations.
We show that all 
dynamical properties of the stream are stationary in action-angle space and
scale with the tidal radius, that is, the 
cube root of the satellite mass,
allowing a single simulation to be adjusted to give the properties of
streams from clusters of lower or higher masses.

The tidal pulses of mass loss lead to a set of particles with strongly correlated motions around the stream, creating a 
spur feature which oscillates above and below the stream centerline, gradually stretching and blurring out. 
With increasing orbital eccentricity the dispersion of angular momentum in the stream rises, increasing
the speed at which particles mix together to smooth out features. 
The stream simulation with an orbital eccentricity of 0.57 is only thin as it passes pericenter and develops a fan of particles in its tail which turns into 
a shell feature at apocenter. 

A specific application of these simulations is to measure the rate at which gaps in a stellar stream blur out as a result of 
both random and differential motions.  The stream with eccentricity 0.57 blurs out features quickly, which means
that we expect gaps as narrow as the stream to survive less than 2 radial orbits in Pal~5 with an eccentricity of 0.6, although
wider gaps will survive longer.
On the other hand, the GD-1 stream, with an eccentricity of 0.33, should allows gaps to survive about 6 radial orbits 
before they blur out. Both of these quantitative conclusions can be refined with a more accurate potential model of the galaxy than 
our simple isochrone.
Overall this paper adds to the understanding of the dynamical makeup of streams and better informs their use
as indicators of the galactic potentials and substructure in those potentials.

\acknowledgements

I thank the referee for constructive comments which led to significant clarifications both in content and presentation of this paper.
This research was supported by CIFAR and NSERC Canada.

\end{document}